# Statistical Voice Conversion with Quasi-Periodic WaveNet Vocoder

*Yi-Chiao Wu[1], Patrick Lumban Tobing[1], Tomoki Hayashi[2], Kazuhiro Kobayashi[3], Tomoki Toda[3]*

[1] Graduate School of Informatics, Nagoya University, Japan
[2] Graduate School of Information Science, Nagoya University, Japan
[3] Information Technology Center, Nagoya University, Japan

`yichiao.wu@g.sp.m.is.nagoya-u.c.jp, tomoki@icts.nagoya-u.c.jp`

## Abstract

In this paper, we investigate the effectiveness of a quasi-periodic WaveNet (QPNet) vocoder combined with a statistical spectral conversion technique for a voice conversion task. The WaveNet (WN) vocoder has been applied as the waveform generation module in many different voice conversion frameworks and achieves significant improvement over conventional vocoders. However, because of the fixed dilated convolution and generic network architecture, the WN vocoder lacks robustness against unseen input features and often requires a huge network size to achieve acceptable speech quality. Such limitations usually lead to performance degradation in the voice conversion task. To overcome this problem, the QPNet vocoder is applied, which includes a pitch-dependent dilated convolution component to enhance the pitch controllability and attain a more compact network than the WN vocoder. In the proposed method, input spectral features are first converted using a framewise deep neural network, and then the QPNet vocoder generates converted speech conditioned on the linearly converted prosodic and transformed spectral features. The experimental results confirm that the QPNet vocoder achieves significantly better performance than the same-size WN vocoder while maintaining comparable speech quality to the double-size WN vocoder.

## 1. Introduction

The main concept of speaker voice conversion (VC) is to convert the speaker identity of a source utterance to a specific target speaker while maintaining the same linguistic contents. A general VC system includes a pipeline for acoustic feature extraction and conversion, and waveform generation based on the converted features. Mainstream VC frameworks focus on the source-target mapping of spectral features that are extracted by traditional source-filter model parametric vocoders such as STRAIGHT [1] and WORLD [2]. There are many approaches to spectral conversion such as the use of the Gaussian mixture model (GMM) [3–5], frequency warping [6, 7], and exemplar-based approaches [8–10]. Furthermore, benefiting from the recent development of deep learning, a variety of neural-based methods have been proposed such as a feedforward deep neural network (DNN) [11–13], variational autoencoder (VAE) [14–16], and recurrent neural network (RNN) [17].

However, because of the oversimplified assumptions in speech signal processing, the traditional parametric vocoder loses some essential information of speech such as the phase. Therefore, conventional vocoder-based VC suffers from serious quality and speaker similarity degradation. To address this issue, many neural-based vocoders [18–23] have been proposed to replace the traditional vocoders in the synthesis part of VC. In this paper, we focus on the WaveNet (WN) vocoder [18–21], which is an autoregressive model conditioned on auxiliary features to generate a raw waveform without many handcrafted assumptions. Although the WN vocoder generate high-fidelity speech conditioned on the training acoustic features, the fixed network architectures of WN are not efficient and may reduce the robustness against unseen fundamental frequency ($F_0$) features that are not observed in the range of training data. Specifically, to achieve acceptable speech quality, the required long *receptive field* of WN results in a huge network size. However, because speech has a quasi-periodic pattern, a fixed long *receptive field* may include many redundant previous samples. As a result, it is more reasonable that each sample has a specific dependent field corresponding to its periodicity. Moreover, the fixed autoregressive structure only implicitly models the relationship between the periodicity of waveform signals and auxiliary $F_0$ values, which may not explicitly generate speech with the correct pitch related to the auxiliary $F_0$ values, especially conditioned on the unseen $F_0$ values.

In our previous work [24], we proposed an augmented quasi-periodic WaveNet (QPNet) vocoder, which included a cascaded structure of several fixed and pitch-dependent (adaptive) dilated convolution layers. The fixed dilated convolution layers modeled the short-term correlation of the current sample and a specific number of previous samples similarly to WN, and the adaptive ones further modeled the long-term correlation related to the conditional $F_0$. The introduced quasi-periodic information gave each sample an exclusive *receptive field* and enhanced the robustness against unseen scaled $F_0$. Moreover, because of the more efficient way of extending the *receptive field* of QPNet, half the network size was required to achieve comparable speech quality to WN.

In this paper, we further investigate the effectiveness of the QPNet vocoder combined with a statistical VC technique. In the proposed system, a framewise DNN model first converts source spectral features to target spectral features, and then the QPNet vocoder generates the converted speech based on the converted spectral and linearly transformed prosodic features. In addition, two speaker adaptation methods for multispeaker WN-based vocoders [25–27] are explored. Both objective and subjective evaluations are conducted, and the experimental results show that the QPNet vocoder with half the network size achieves significantly better performance than the WN vocoder with the same size while maintaining comparable speech quality and speaker similarity to the full-size WN.

## 2. Related work

WaveNet [18] as one of the state-of-the-art audio generation models has been widely applied to various VC systems that take

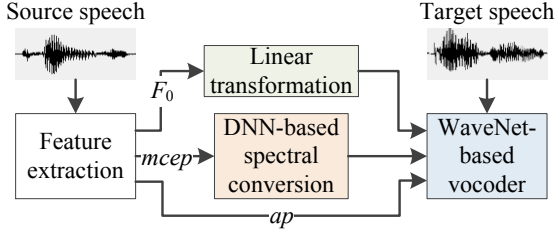

Figure 1: *DNN-based VC with WaveNet-based vocoder*

WN as a vocoder to generate converted waveforms from the converted acoustic features. For example, Kobayashi et al. [28] combined GMM-based Mel-cepstral coefficient (*mcep*) conversion and linear transformation of prosodic features with the WN vocoder. Furthermore, in our previous works, we explored the effectiveness of different *mcep* conversion models with the WN vocoder, including a DNN [25, 29], deep mixture density network (DMDN) [26], VAE [30], long short-term memory (LSTM) [31], and gated recurrent unit (GRU) [32]. Inspired by Tacotron2 [33], Chen et al. [34] and Zhang et al. [35] proposed conditioning WN on a Mel-spectrogram to obtain better speech quality than *mcep*-based methods. Benefiting from the success of VC with the advances in extra automatic speech recognition (ASR) systems [36, 37], Liu et al. [38] and Sisman et al. [27] also proposed different VC models with the aid of a phonetic posteriorgram (PPG) combined with the WN vocoder. On the other hand, Niwa et al. [39] and Tian et al. [40] also proposed direct WN-based VC without additional feature conversion models. However, different from these previous studies in this paper, we focus on improving the WN vocoder for VC rather than enhancing the spectral conversion accuracy.

## 3. Baseline voice conversion system with WaveNet-based vocoder

The general flowchart of DNN-VC with a WN-based vocoder is shown in Fig. 1, which includes acoustic feature extraction by the conventional parametric vocoder, DNN-based source-target feature conversion, and WN-based converted waveform generation conditioned on the converted acoustic features. In this section, we describe the DNN-VC and WN vocoder modules of the baseline VC system.

### 3.1. DNN-based spectral conversion

DNN-based spectral conversion [41, 42] includes training and conversion stages. Specifically, the neural network models the relationship between the given source static-dynamic feature vector $\mathbf{S}_n = [\mathbf{s}_n^T, \Delta \mathbf{s}_n^T]^T$ and the target static-dynamic feature vector $\mathbf{T}_n = [\mathbf{t}_n^T, \Delta \mathbf{t}_n^T]^T$ at frame $n$ using the conditional probability density function as follows:

$$P(\mathbf{T}_n | \mathbf{S}_n, \mathbf{\Sigma}, \mathbf{\lambda}) = \mathrm{N}(\mathbf{T}_n; \mathrm{f}_{\lambda}(\mathbf{S}_n), \mathbf{\Sigma}), \quad (1)$$

where $\mathrm{N}(\cdot)$ denotes a Gaussian distribution, $\mathrm{f}_{\lambda}(\cdot)$ is the nonlinear conversion function of the DNN, $\mathbf{\lambda}$ represents the DNN parameters, and $\mathbf{\Sigma}$ is the diagonal covariance matrix of the training data. In the training stage, the updated form of the DNN parameters $\hat{\mathbf{\lambda}}$ is as follows:

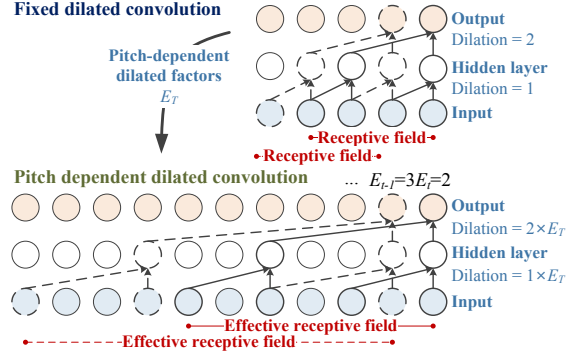

Figure 2: *Pitch-dependent dilated convolution*

$$\hat{\mathbf{\lambda}} = \underset{\lambda}{\mathrm{argmax}} \sum_{n=1}^{N} \log P(\mathbf{T}_n | \mathbf{S}_n, \mathbf{\Sigma}, \mathbf{\lambda})$$
$$= \underset{\lambda}{\mathrm{argmin}} \frac{1}{2} \sum_{n=1}^{N} (\mathbf{T}_n - \mathrm{f}_{\lambda}(\mathbf{S}_n))^T \mathbf{\Sigma}^{-1} (\mathbf{T}_n - \mathrm{f}_{\lambda}(\mathbf{S}_n)). \quad (2)$$

In the conversion stage, given the DNN output, the trajectory of the target feature vector is generated by maximum likelihood parameter generation (MLPG) [43]. Furthermore, to minimize the oversmoothing effect caused by the averaging in the model, a global variance (GV) [5] postfilter is applied.

### 3.2. WaveNet vocoder

To model the very long term dependence of speech signals, WN predicts the conditional distribution of the current speech sample with input auxiliary features and a specific number of previous samples, which is called the *receptive field*. The conditional probability can be formulated as

$$P(\mathbf{Y} | \mathbf{h}) = \prod_{t=1}^{T} P(y_t | y_{t-1}, ..., y_{t-r}, \mathbf{h}), \quad (3)$$

where $t$ is the sample index, $r$ is the length of the *receptive field*, $y_t$ is the current audio sample, and $\mathbf{h}$ is the vector of the auxiliary features. Moreover, because of the causality of speech signals and the efficiency of extending *receptive field* for the long-term correlation of speech, WN applies a stacked dilated causal convolution structure [18, 44]. Moreover, the following gated structure is applied to enhance the modeling capability:

$$\mathbf{Z} = \tanh(\mathbf{V}_{f,k}^{(1)} * \mathbf{Y} + \mathbf{V}_{f,k}^{(2)} * u(\mathbf{h})) \otimes \sigma(\mathbf{V}_{g,k}^{(1)} * \mathbf{Y} + \mathbf{V}_{g,k}^{(2)} * u(\mathbf{h})), \quad (4)$$

where $\mathbf{V}^{(1)}$ and $\mathbf{V}^{(2)}$ are trainable convolution filters, $*$ is the convolution operator, $\otimes$ is an elementwise multiplication operator, $\sigma$ is a sigmoid function, $k$ is the layer index, $f$ and $g$ represent the filter and gate, respectively, and $u(\cdot)$ is an upsampling layer used to adjust the resolution of auxiliary features to match that of input speech samples. Moreover, 8-bit $\mu$-law encoding is applied to the output waveform of WN, which makes the WN output become a categorical distribution. For the WN vocoder, the previous speech samples pass through a pipeline including a causal layer and several residual blocks which contains a dilated convolution layer, gated activation with auxiliary acoustic features, and residual and skip connections. Then, the summation of all skip connections is passed to two 1×1 convolution and one softmax layers to output the predicted distribution of the current sample.

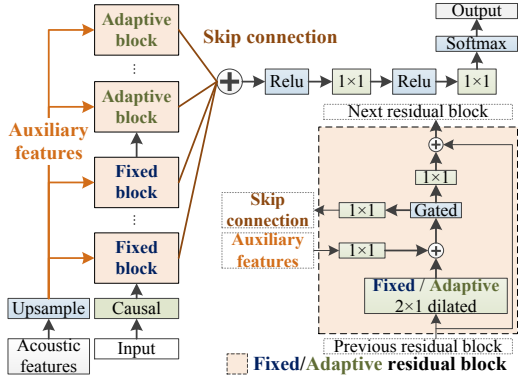

Figure 3: *Quasi-periodic WaveNet vocoder architecture*

| Hyperparameter | WNf | WNc | QPNet |
|---|---|---|---|
| Number of fixed layers | 10 | 4 | 4 |
| Number of fixed repeats | 3 | 4 | 3 |
| Number of adaptive layers | - | - | 4 |
| Number of adaptive repeats | - | - | 1 |
| Constant $a$ | - | - | 8 |
| Causal and dilated conv. | 512 channels | | |
| 1×1 conv. in residual blocks | 512 channels | | |
| 1×1 conv. between skip-connection and softmax | 256 channels | | |

Table 1: *Comparison of hyperparameters*

## 4. Proposed voice conversion system with Quasi-Periodic WaveNet vocoder

In this section, the advanced QPNet vocoders and the proposed VC system are introduced as follows.

### 4.1. Quasi-periodic WaveNet vocoder

The main differences between the quasi-periodic WN and the vanilla one are in the pitch-dependent dilated convolution and cascade network structure. Specifically, the pitch-robust dilated convolution, which is inspired by the pitch filtering in the code-excited linear prediction (CELP) codec [45], makes the size of the *receptive field* become pitch-related by dynamically changing the dilation size of the convolution according to the $F_0$ values of the input signals, whereas the *receptive field* size of vanilla WN is time-invariant. To elaborate this concept as shown in Fig. 2, the dilated convolution can be formulated as

$$\mathbf{X}_t^{(o)} = \mathbf{W}^{(c)} * \mathbf{X}_t^{(i)} + \mathbf{W}^{(p)} * \mathbf{X}_{t-d}^{(i)}, \quad (5)$$

where $\mathbf{X}^{(i)}$ is the input and $\mathbf{X}^{(o)}$ is the output of the dilated convolution layer. $\mathbf{W}^{(c)}$ and $\mathbf{W}^{(p)}$ are trainable 1×1 convolution filters of the current and previous samples, respectively. The dilation size $d$ is time-variant and related to the pitch in the pitch-dependent dilated convolution, whereas the vanilla dilated convolution has a constant $d$. Therefore, each convolution layer with the pitch-dependent dilation size models the relationship between the current sample and the relevant sample of multiple previous frequency periods, and it makes the network efficiently extend the *receptive field* without losing trajectory information of the sequential signals.

Furthermore, since speech is a quasi-periodic signal, QPNet respectively models the periodic and nonperiodic components of speech with the adaptive (pitch-dependent) and fixed modules. Specifically, the adaptive module models the long-term periodic correlations of the periodic parts with the given pitches, and the fixed module estimates the short-term information of the nonperiodic parts using the nearest samples. As shown in Fig. 3, the fixed module of the QPNet vocoder is the same as that of the vanilla WN vocoder, which has a causal layer and several stacked residual blocks including fixed dilated convolutions, conditional auxiliary features, gated activations, and residual and skip connections. The adaptive module also has several similar stacked residual blocks but with the fixed dilated convolutions replaced with the pitch-dependent ones.

In addition, the dilation size of each pitch-dependent dilated convolution layer is double that in the previous layer up to a specific number and then repeated, which is consistent with the fixed layers but multiplied by an extra dilated factor related to the $F_0$ values. The sequence of the pitch-dependent dilated factors $E_t$ is as follows:

$$E_t = F_s / (F_{0,t} \times a), \quad (6)$$

where $F_s$ is the sampling rate and $F_0$ is the fundamental frequency of the speech sample, $a$ is a hyperparameter, and $t$ is the sample index. On the basis of this mechanism, each speech sample has the specific length of the *receptive field* matched to its pitch. Moreover, $a$ is the number of samples in one cycle that are taken into consideration by the network to predict the next sample, which we empirically set to 8 in this paper. To ensure the speech quality, the interpolated continuous $F_0$ values are adopted to obtain the pitch-dependent dilated factors.

### 4.2. Implementation for voice conversion

The proposed framework includes training and testing stages. All speech data are first processed by the WORLD vocoder to extract the spectral (*sp*), $F_0$, and aperiodic (*ap*) features, and *sp* is further parameterized into *mcep*. In the training stage, the multispeaker QPNet vocoder and speaker-pair-dependent DNN-based spectral VC models are trained with the training corpus, and then the target-speaker-dependent QPNet vocoders are further updated from the multispeaker QPNet vocoder using every target speaker's training data. In the testing stage, the source *mcep* is converted to a specific target by the trained DNN-VC model, and then the speaker-dependent QPNet vocoder generates the converted speech waveforms conditioned on the converted *mcep*, linearly transformed $F_0$, and source *ap*.

## 5. Experimental evaluations

In this section, we first investigate the updating strategy of speaker-dependent (SD) WN-based vocoders, which include a full-size WN vocoder (WNf), a compact-size WN vocoder (WNc), and a compact-size QPNet vocoder. Furthermore, we conducted objective tests to evaluate the waveform generation capability of the vocoders and subjective tests to evaluate the performance of the proposed VC system.

### 5.1. Experimental settings

The training corpus of the multispeaker WN-based vocoders included the training data of the 'bdl' and 'slt' speakers of CMU-ARCTIC [46] and all the training data of VCC2018 [47]. The hyperparameters of the network structure are shown in Table 1. The training procedure was consistent with [26]. The testing set of each VCC2018 speaker contained 35 utterances, and we further divided them into five utterances for validation and 30 utterances for VC performance evaluations. In addition, we took the four speakers (two males and two females) of the VCC2018 SPOKE set as the source speakers, and formed 16 VC speaker pairs with the four target speakers (two males and two females) of the VCC2018 HUB set. The DNN-VC models were trained with the training set of the corresponding speaker pairs, which included 81 utterances of each speaker, and extra reference utterances generated by a text-to-speech (TTS) system [26]. The feedforward DNN models included four hidden layers with 1024 hidden units per layer, and the training also followed our previous system [26] submitted to VCC2018.

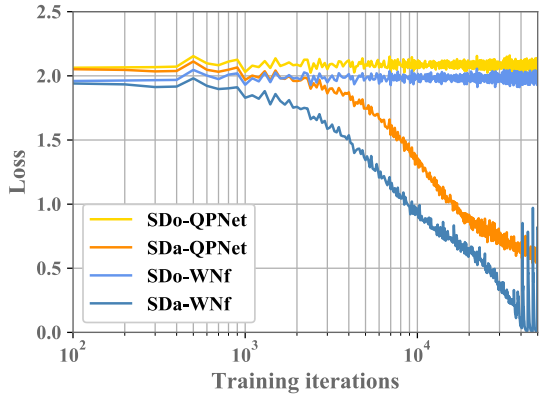

Figure 4: *Adaption training losses of different vocoders* (SD: *speaker-dependent;* o: *only update output layers;* a: *update whole network*)

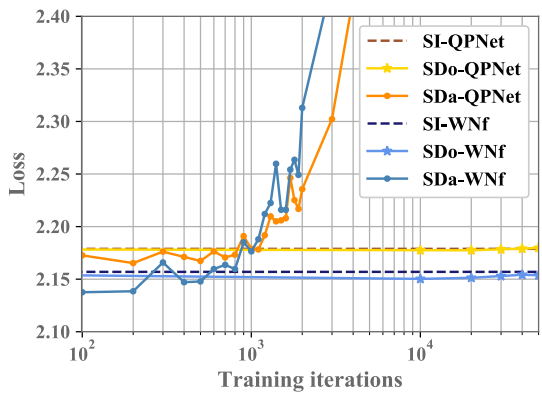

Figure 5: *Validation losses of different vocoders* (SI: *speaker-independent;* SD: *speaker-dependent;* o: *only update output layers;* a: *update whole network*)

Table 2: *Comparison of mel-cepstral distortion and root-mean-square error of log $F_0$ with different vocoders*

|  |  | WNf | WNc | QPNet |
|---|---|---|---|---|
| MCD | SI | 3.25 | 3.83 | 3.57 |
|  | SDo | 3.11 | 3.73 | 3.51 |
|  | SDa | **3.02** | 3.68 | 3.46 |
| RMSE of log $F_0$ | SI | 0.15 | 0.21 | 0.15 |
|  | SDo | 0.15 | 0.20 | **0.13** |
|  | SDa | 0.15 | 0.19 | 0.14 |

All speech data were set to sampling rate 22050 Hz and 16 bits resolution. One-dimensional $F_0$ and 513-dimensional *sp* and *ap* features were extracted by WORLD, and *sp* is further parameterized into 34-dimensional *mcep*. For the WN-based vocoder, $F_0$ values were converted into continuous $F_0$ features and voice/unvoice (*uv*) binary symbols, *ap* features were coded into two-dimensional components [19], and speech waveforms were encoded into 8 bits by the $\mu$-law. For VC, the source *mcep* was conveted by a DNN-VC model, the source $F_0$ was linearly transformed in the logarithm domain, and the source *ap* was directly adopted for the converted acoustic features.

### 5.2. Speaker-dependent WaveNet adaptation

In this section, we survey two fine-tuning strategies to update the speaker-independent (SI) WN-based vocoders to SD ones, which involve updating all network parameters (SDa) and only updating the final output layers of the network (SDo) with the training data of the target speakers. Figure 4 shows the training loss (cross-entropy) of the SD vocoders with speaker TM1, although the other target speakers had the same tendency. The utterances used for fine-tuning were only the 81 utterances of TM1, the updating batch size was 20,000 samples, and the number of iterations were from 100 to 50,000. As shown in Fig. 4, the training losses of the SDa vocoders started to decrease remarkably when beyond 1000 iterations, whereas the training losses of the SDo vocoders were stable regardless of the number of iterations, which might indicate that updating the whole network with very limited data will cause serious overfitting. In addition, we measured the fine-tuning performance using the training loss of the validation data while fixing the network parameters denoting the validation loss. Figure 5 shows that the validation losses of the SDa vocoders started to increase from around 500 iterations (~2 epochs), whereas the SDo vocoders exhibited stable validation losses. Therefore, we set the number of updating iteration as 500 for the SDa vocoders and 50,000 for the SDo vocoders (our system submitted to VCC2018 [26] was SDo-WNf with 50,000 iterations) in this paper. In the next section, we further evaluated the generation capability of these vocoders with VC features.

### 5.3. Objective evaluations

To evaluate the converted waveform generation performance of the vocoders with statistically converted acoustic features, we measured the Mel-cepstral distortion (MCD) and the root-mean-square error (RMSE) of logarithmic $F_0$ between the acoustic features extracted from the converted speech and the auxiliary features of the vocoders. Specifically, we computed MCD between the conditional and extracted *mcep* to evaluate the robustness of spectrum reconstruction with the vocoders conditioned on the VC acoustic features. Moreover, to evaluate the generation pitch accuracy of each vocoder corresponding to the conditional linearly transformed $F_0$, we calculated the RMSE between the conditional $F_0$ and the $F_0$ extracted from the converted speech.

As shown in Table 2, the QPNet vocoder significantly outperformed the WNc vocoder with the same network size as QPNet in both MCD and RMSE measurements. Even compared with the WNf vocoder with double the network size, the QPNet vocoder still achieved slightly higher pitch generation accuracy. Although the WNf vocoder had the highest spectrum prediction capability, QPNet still outperformed WNc. The much shorter *receptive field* caused by the halved network size might degrade the spectral prediction capability of QPNet. In summary, the objective evaluations show that the pitch-dependent dilation structure of QPNet can increase the capability of spectrum prediction and the accuracy of the pitch for WN-based vocoders.

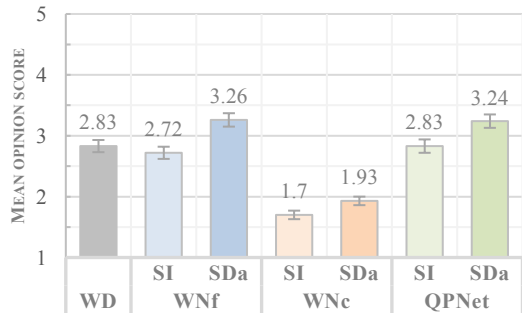

Figure 6: *MOS evaluation of sound quality with 95% confidence intervals*. (WD: *WORLD vocoder*; SI: *speaker-independent vocoder*; SDa: *speaker-dependent vocoder with fine-tuning of whole network*)

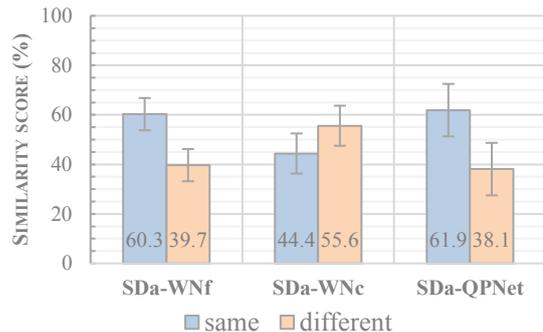

Figure 7: *Speaker similarity evaluation with 95% confidence intervals*. (SDa: *speaker-dependent vocoder with fine-tuning of whole network*)

Furthermore, all SDo and SDa vocoders achieved better MCD than the relevant SI vocoders, and the results confirmed the effectiveness of the SD fine-tuning. Because the SDa vocoders attained the highest spectrum prediction capabilities, we applied the updating strategy of SDa to the SD-VC systems in this paper. In the next section, we conducted subjective tests to evaluate the VC quality of the waveforms from the different vocoders.

### 5.4. Subjective evaluations

To evaluate the speech quality and speaker similarity of the converted waveforms generated by the different vocoders with the converted acoustic features, we conducted mean opinion score (MOS) and speaker similarity tests. Specifically, we randomly selected 20 utterances from 30 testing utterances of each speaker pair and vocoder to establish an evaluation set. Then, we divided the set into 10 non-overlapping subsets for 10 listeners, and each subset was evaluated by one listener. As a result, each listener evaluated 224 different utterances generated by seven vocoders including the SI and SDa WN-based vocoders and WORLD in the MOS test. The speech quality was assigned a value of 1–5; the higher the score, the better the naturalness. Moreover, the speaker similarity evaluation followed the test flow of VCC2018 [47]. That is, a subject was first asked to listen to a natural speech and a converted speech, and then asked to evaluate the speaker similarity of the two speech files using four labels: *definitely the same*, *maybe the same*, *maybe different*, and *definitely different*. The final speaker similarity scores are the sum of the percentages of *definitely the same* and *maybe the same* and the sum of *definitely different* and *maybe different*.

As shown in Fig. 6, the MOS evaluation results of WNc and QPNet indicate that the pitch-dependent dilated convolution significantly improved the speech quality of converted speech even though the network sizes of the two vocoders were the same. Furthermore, the overall results confirmed the effectiveness of the SD fine-tuning of all WN-based vocoders to achieve significantly better speech naturalness. Compared with the full-size WN vocoder, SI-QPNet attained slightly better performance than SI-WNf, and the perceptual qualities of SDa-QPNet and SDa-WNf were comparable despite the network size of QPNet being only half of that of WNf. Moreover, SDa-QPNet also achieved demonstrably better conversion speech generation capability than the traditional WORLD vocoder. To further evaluate the conversion accuracy of speaker identity among the WN-based vocoders, we conducted the speaker similarity tests on the SDa-WNf, SDa-WNc, and SDa-QPNet vocoders. The results in Fig. 7 demonstrate the same tendency as the speech naturalness results. SDa-QPNet markedly outperformed SDa-WNc for speaker similarity and achieved similar performance to SDa-WNf. The demo can be found at "https://bigpon.github.io/QuasiPeriodicWaveNet_demo/".

## 6. Conclusions

In this paper, we investigated the speaker conversion speech generation performances of the QPNet vocoder compared with the full- and compact-sized WN vocoders and the traditional WORLD vocoder. The inputs of each vocoder are the spectral features converted by a framewise DNN-VC model and linear-transformed prosodic features. Furthermore, we also evaluated the effectiveness of two speaker adaption methods for SD WN-based vocoders. Both objective and subjective evaluations confirmed the effectiveness of the speaker adaption technique and the QPNet vocoder, which takes advantage of the pitch-dependent dilated convolution to attain better pitch controllability and achieve comparable quality to the WN vocoder with only half the network size. In future works, we will survey different combinations of the pitch-dependent and fixed dilated convolutions to achieve optimized performance.

## 7. Acknowledgements

This work was partly supported by JST, PRESTO Grant Number JPMJPR1657, and JSPS KAKENHI Grant Number 17H01763.